\documentstyle{article}
\begin{document}
\baselineskip=12pt
\baselineskip=14pt
\title{\bf A Study of Two-Temperature Non-Equilibrium Ising Models:
Critical Behavior and Universality}

\author{P. Tamayo\\
	{\em Theoretical Division}\\
	{\em Los Alamos National Laboratory MS-B285}\\
	{\em Los Alamos, NM 87545}\\
	{\em and}\\
	{\em Thinking Machines Corp.}\\
	{\em 245 First St.}\\
	{\em Cambridge, MA 02142}\\[.4cm]
\and
	F. J. Alexander\thanks{Present address: Institute 
	for Scientific Computing Research L-416, 
        Lawrence Livermore National Laboratory,
	Livermore, CA 94550}\\
        {\em Center for Nonlinear Studies}\\
        {\em Los Alamos National Laboratory MS-B258}\\
        {\em Los Alamos, NM 87545}\\[.4cm]
\and
	R. Gupta\\
	{\em Theoretical Division}\\
	{\em Los Alamos National Laboratory MS-B285}\\
	{\em Los Alamos, NM 87545}\\[.4cm]}

\date{}
\maketitle
\vspace{0.4in} 

\begin{center}
{\bf ABSTRACT}
\end{center}

We  study a  class  of  $2D$  non-equilibrium  Ising  models based  on
competing dynamics induced by contact with heat-baths at two different
temperatures.   We make a comparative  study  of  the  non-equilibrium
versions  of Metropolis, heat bath/Glauber  and Swendsen-Wang dynamics
and  focus on their  critical behavior  in  order to  understand their
universality classes.  We  present strong evidence  that some of these
dynamics  have the  same  critical exponents  and  belong  to the same
universality class as the equilibrium $2D$ Ising  model.  We show that
the  bond version of the Swendsen-Wang update algorithm can  be mapped
into an equilibrium model at an effective temperature.

\newpage

{\Large \bf 1.- Introduction}
\bigskip

Despite  attempts   at  constructing   a   rigorous  theory  for  {\em
non-equilibrium} statistical mechanics, there is still no formalism to
parallel  the one  which  exists for  {\em equilibrium} systems.  As a
result,  there are  few  analytical methods with  which  to  deal with
non-equilibrium  systems. In general, non-equilibrium  systems  display
rich and complex behavior such as phase separation, pattern formation,
turbulence~\cite{GUN,CR},  and it  is  therefore useful to first study
simple model systems.

There  exist model  non-equilibrium  systems described  by  stationary
distributions  that  are comparatively easy  to study.  An open system
maintained   in  a  non-equilibrium   steady   state  by  an  external
temperature  or  density  gradient  is   one  example~\cite{SPOHN}.
Another  class of non-equilibrium  steady states is obtained when  the
system  is closed,  and the  dynamics is  a local  competition  of two
dynamics at different temperatures~\cite{GLM}.  It is this type of
system that  we address in  this paper.   Since  these non-equilibrium
systems display behavior qualitatively similar to equilibrium systems,
such as phase transitions, it  is important to ask if  their  critical
properties  and  universality   classes  are  the   same.   Flows   of
renormalized probability  distributions  and fixed points are normally
independent of the details of a Hamiltonian.   Perhaps these flows and
fixed points are even  independent of the  existence of a  Hamiltonian
and  a  Boltzmannian distribution.  If  one  were  able  to  establish
equivalences (or near  equivalences)  of  universality classes between
equilibrium and non-equilibrium models, then we will be able to answer
many questions about the behavior of these systems without  a complete
formulation equivalent to the one for equilibrium systems.

Simple  non-equilibrium  spin-flip  stochastic  systems  have received
considerable  attention in  the  literature over  the  last 10  years.
Reviews   and   general   discussions  about   non-equilibrium   phase
transitions    and    stationary    states    can    be    found    in
refs.~\cite{MARRO,LGM,MVG,GM}.  Studies  of driven  diffusive  systems
can  be  found  in   Wang  {\it  et  al}~\cite{WBL},  Marro  {\it   et
al}~\cite{MGV}, Garrido {\it et  al}~\cite{GMD}  and Grinstein {\it et
al}~\cite{GJS}.

Grinstein {\it et al}~\cite{GJH} studied the  statistical mechanics of
probabilistic cellular automata  using  time-dependent Ginzburg-Landau
theory.  They  suggested  that any  non-equilibrium spin-flip dynamics
with up-down symmetry belongs to the same universality  class  as  the
equilibrium Ising model.  Their  argument is based on the  observation
that  under  the  renormalization  group  in  $d  =  4-\epsilon$,  the
dynamical fixed point of the Ising model is stable with respect to all
additional analytic terms which preserve the lattice geometry and  the
spin up-down symmetry.

Kanter  and Fisher~\cite{KF} analyzed the  existence of ordered phases
in  stochastic Ising systems with short-range interactions. They found
that the existence of universality for those  systems might depend  on
the  details  of  the interaction and  casted  doubts on  the  general
applicability of the argument of Grinstein {\it et al}~\cite{GJH}.

A  two-temperature Glauber  Ising  model  was  introduced by  Garrido,
Labarta and Marro~\cite{GLM} (we will refer to this  model as GLM)  to
investigate stationary  non-equilibrium states.  They  obtained a mean
field  solution  and performed  some Monte  Carlo  simulations on $2D$
lattices.  They found  critical behavior qualitatively  similar to the
equilibrium case.

The non-equilibrium behavior of competing dynamics such  as  spin-flip
{\it   vs}   spin-exchange  has  been  studied  by  Garrido  {\it   et
al}~\cite{GMG}  using  hydrodynamic  macroscopic  equations and  Monte
Carlo data, and by  Wang  and Lebowitz~\cite{WL}  using a  Monte Carlo
renormalization group method. They found evidence of equilibrium Ising
behavior and Ising-like exponents.

Tom\'e {\it et al}~\cite{TOS}  studied the GLM model for the case when
one of the  temperatures is  negative.   They  used a  dynamical  pair
approximation to analyze  antiferromagnetic steady states and obtained
the  corresponding phase  diagram.   Their mean  field renormalization
group calculations show evidence in favor of equivalence with  the
equilibrium Ising universality class.

Bl\"ote {\it et  al}~\cite{BHHZ} studied  a model  similar  to GLM  in
which each  sublattice is  in equilibrium  at a different temperature.
They performed Monte Carlo simulations  and found strong evidence that
the model belongs to the equilibrium Ising universality class.

Marques~\cite{MAR-I,MAR-II}  used a mean field  renormalization  group
calculation to obtain  a  phase diagram  and calculated  the  exponent
$\nu$  for  the  GLM  model.   Her  results  compared  well  with  the
equilibrium  Ising  values.   Later she extended this technique to two
different   three   state    systems   which   retain    the   up-down
symmetry~\cite{MAR-III}   to    test    Grinstein   {\it    et    al.}
conjecture~\cite{GJH}, and found good agreement with equilibrium Ising
exponents.

Recently  de Oliveira~\cite{OLI}  analyzed the isotropic majority-vote
non-equilibrium model by Monte Carlo and finite size scaling. He found
very good agreement for the critical  exponents between this model and
the  equilibrium  Ising  model, and also  for Binder's cumulant.  In a
separate paper de Oliveira  {\it et al}~\cite{OMS} studied a family of
non-equilibrium spin models  with up-down symmetry parameterized  by a
Glauber-like transition  rate.  They also found good evidence that the
critical exponents for this family  (except  for the limit case of the
voter model) are the same as the equilibrium Ising model.

Considerable numerical and analytic evidence  has been accumulating in
favor  of universality and  equilibrium Ising exponents  for  some  of
these models  but the results are not definite yet.   Most  analytical
methods used,  such as  mean-field RG, are  of an approximate  nature.
Monte Carlo  simulations  have  focused mainly on qualitative behavior
and the calculations of critical exponents have  not been carried  out
with  high resolution.  To better understand and test this equivalence
we have  undertaken  a  detailed, high resolution Monte Carlo study of
two-temperature Ising models and  explore the question of universality
using different local and non-local update dynamics.

The  paper is organized as  follows. In Section 2 we will describe the
different  two-temperature non-equilibrium models that are the subject
of this study.  Section 3 presents detailed results for the Metropolis
non-equilibrium  dynamics  including  critical exponents and  cumulant
behavior.    Section   4   focuses   on  the   Swendsen-Wang~\cite{SW}
non-equilibrium  dynamics. Section  5 contains  a comparative study of
all the dynamics.  Extension to many temperature model is discussed in
Section 6 and the conclusions are presented in Section 7.

\bigskip
\bigskip
{\Large \bf 2. Two-temperature non-equilibrium dynamics}
\bigskip

We begin with the two dimensional Ising model on the square lattice
with Hamiltonian
\begin{equation}
H = - J\beta \sum_{\langle ij \rangle} \sigma_i \sigma_j
\end{equation}
where $\beta$ is the inverse temperature and $J$ the coupling.  A
dynamics for the model can be described in terms of a time dependent
probability distribution $P(\sigma, t)$ which evolves according to a master
equation,
\begin{equation}
{dP(\sigma, t) \over dt} = \sum_{\sigma'} \{ W(\sigma' \leftarrow \sigma)
P(\sigma, t) -  W(\sigma \leftarrow \sigma') P(\sigma', t) \}
\end{equation}
where $W(\sigma' \leftarrow \sigma)$ is the transition rate from
configuration $\sigma$ to $\sigma'$. We are interested in stationary
probability distributions,
\begin{equation}
{dP(\sigma, t) \over dt} = 0, \;\;\; P(\sigma, t) = P(\sigma)
\end{equation}
In the case of equilibrium systems $P(\sigma)$ is not only stationary but
has the form of a Boltzmann distribution parameterized by the inverse
temperature $\beta$,
\begin{equation}
P(\sigma) = Z^{-1} \exp(- \beta H)
\end{equation}
where $Z=\sum_{\sigma} \exp(-\beta  H (\sigma ))$.  In our case we are
interested in stationary non-equilibrium distributions produced by the
local competition of equilibrium  dynamics  at different temperatures.
The usual condition  to obtain an equilibrium  Monte Carlo dynamics is
to make $W(\sigma'  \leftarrow  \sigma)$ obey detailed  balance  (note
that  imposing detailed  balance on the  $W$ is a  sufficient but  not
necessary condition),
\begin{equation}
 W(\sigma' \leftarrow \sigma) P(\sigma) =  W(\sigma \leftarrow \sigma') 
 P(\sigma') .
\end{equation}
In the two-temperature model one considers a composite rate
$W(\sigma'
\leftarrow \sigma)$,
\begin{equation}
W(\sigma' \leftarrow \sigma) =  p W_1(\sigma' \leftarrow \sigma)
 +  (1-p) W_2(\sigma' \leftarrow \sigma) 
\end{equation}
with competing $W_1$ and $W_2$. At each time step the transition
probability will be chosen at random to be $W_1$ with probability $p$ or
$W_2$ with probability $(1 - p)$. $W_1$ and $W_2$ individually
correspond to equilibrium transition rates obeying detailed balance with
respect to temperatures $\beta_1$ and $\beta_2$, $i.e.$
\begin{equation}
{W_i(\sigma \leftarrow \sigma') \over W_i(\sigma' \leftarrow \sigma)} =
{e^{\beta_i H (\sigma)} \over  e^{\beta_i H (\sigma')}} .
\end{equation}
At  each time  step the dynamics obeys  detailed balance locally  with
respect  to  $\beta_1$  or  $\beta_2$  and  the  spins  act  as if  in
instantaneous contact with one of two heat baths.   The overall effect
is  to  produce  a  non-equilibrium  dynamics  which  reduces  to  the
equilibrium model when $\beta_1 = \beta_2$.  From  the master equation
one can prove  that  for a combined  dynamics  of this  sort there  is
indeed a stationary regime given by the condition~\cite{MARRO},
\begin{equation}
p \langle \sigma W_1 \rangle + (1 - p) \langle \sigma W_2 \rangle = 0
\end{equation}
The induced global probability  distribution $P(\sigma)$  does not, in
general, correspond to a local known Hamiltonian and it depends on the
details  of the dynamics, $i.e.$ the  particular  choice of  $W_1$ and
$W_2$.  This  is  in contrast  to  equilibrium simulations  where  the
Boltzmann distribution is independent of the particular choice of $W$.
The  combination  of  $W_1$ and  $W_2$  produces  a  stationary  state
analogous  to a  system being driven by  an external  potential.   The
ensemble  of  stationary configurations exhibits  physical  properties
qualitatively  similar   to   equilibrium  (i.e.    ordering,  cluster
formation, phase transitions, etc.).   This is  therefore  one of  the
simplest  ways  to  generate  non-equilibrium models from  equilibrium
ones.   In  order   to   investigate  the   properties  of  stationary
distribution  on  the   dynamics  we  study  three  different   update
algorithms:  Metropolis,   Glauber  (or  equivalently  heat-bath)  and
Swendsen-Wang.  Furthermore each of these dynamics has  a ``bond'' and
``spin'' version as we explain below.

We start by defining the GLM dynamics~\cite{GLM}.  The transition rate
for spin $i$ takes the standard Glauber form,
\begin{equation}
W_i = {1 \over 2} \alpha [ 1 - \tanh(J \beta_i \sigma_i \sum_{|j - i| = 1} \sigma_j) ]
\end{equation}
where $\beta_i$ is chosen  to be equal  to $\beta_1$  with probability
$p$  and  $\beta_2$ with probability $(1 - p)$.  In one dimension this
dynamics is always equivalent to an equilibrium  model at an effective
temperature $\beta_{\rm eff}$ given by~\cite{GLM},
\begin{equation} 
\tanh(2 J \beta_{\rm eff}) = p \tanh(2 J \beta_1) + (1 - p) \tanh(2 J \beta_2)
\end{equation}
Similarly ``bond'' dynamics is defined by 
\begin{equation}
W_i = {1 \over 2} \alpha [ 1 - \tanh(J  \sigma_i \sum_{|j - i| = 1} \beta_j \sigma_j) ]
\end{equation}
where $\beta_j$ is $\beta_1$  with probability  $p$ or $\beta_2$  with
probability  $(1 -  p)$  and in this  way  the temperature is selected
independently for each bond.  These transition rates with $\alpha = 1$
also  define the heat-bath  algorithm, $i.e.$  the two algorithms  are
equivalent and we only  need to discuss one.  Henceforth we shall refer 
to this as the Glauber dynamics.

Metropolis non-equilibrium dynamics are defined in a similar way. The
relevant acceptance factor for spin $i$ is given by,
\begin{equation}
A_i = e^{- 2 J \sigma_i \beta \sum_j \sigma_j}
\end{equation}
where, as in the Glauber case, $\beta$ is either $\beta_1$ or $\beta_2$.
In the bond version we have,
\begin{equation}
A_i = e^{- 2 J \sigma_i \sum_j \beta_j \sigma_j}
\end{equation}
where each bond $j$ is chosen independently with temperature $\beta_1$ or
$\beta_2$.

Finally  we  define  non-equilibrium Swendsen-Wang~\cite{SW} spin  and
bond  versions. In the bond  version  the percolation probability  for
bond $ij$ is chosen to be,
\begin{equation}
\pi_{ij} = 1 - e^{- 2 J \beta_{ij}}
\end{equation}
where $\beta_{ij}$ is $\beta_1$ with probability $p$ or $\beta_2$ with
probability  $(1-p)$.  The subsequent percolation, cluster finding and
flipping  steps  are  done  in  the  same  way  as  in  the   original
Swendsen-Wang dynamics. In section 4 we show that this dynamics can be
mapped  to  an  equilibrium  system  at  an intermediate ``effective''
temperature $\beta_{\rm eff}$.  The spin version is  defined similarly
except that the four bonds contributing  to the update of each red (or
black) site are  chosen  to  be  at the same  temperature $\beta_1$ or
$\beta_2$.

All these algorithms can be implemented very efficiently on a parallel
computer such as the CM~-~5.  In  this paper we present detailed results
for  the Metropolis-spin  and  Swendsen-Wang-bond cases  and make some
comparisons with the other dynamics.

\bigskip
\bigskip
{\Large \bf 3. Metropolis non-equilibrium dynamics}
\bigskip

We  performed  a  careful investigation of  the  spin  version  of the
Metropolis  dynamics.  Our  update  algorithm  is   parallel,   so  we
simultaneously update  all  the red/black  sites on the (checkerboard)
lattice. For  each sublattice site $i$ we choose temperature $\beta_1$
or  $\beta_2$  independently  using  a  uniformly  distributed  random
number. Then we compute the change of energy with the spin flipped,
\begin{equation}
\Delta E_i = 2 \sigma_i \sum_j \sigma_j.
\end{equation}
If $\Delta E_i$ is less or equal to zero then the flip is always accepted,
otherwise it is accepted with probability $\exp(- \beta_i \Delta E_i)$.

To address the question of the existence of an equivalent equilibrium
system we consider the local transition rate for the combined dynamics,
\begin{equation}
e^{-\beta_{\rm eff} \Delta E} = p e^{-\beta_1 \Delta E} + (1-p) e^{-\beta_2 \Delta E} .
\end{equation}
This equation  has  to  be  satisfied  for all values  of $\Delta  E$,
however,  only  the  cases   for  $\Delta  E  >   0$  are  temperature
dependent\footnote{We  thank  R.~Swendsen  for bringing  this  to  our
attention}.  In one-dimension there is only one relevant case ($\Delta
E = 4$) and the equation is  satisfied with a $\beta_{\rm eff}$ equal to
\begin{equation}
\beta_{\rm eff} = -{1 \over 4} \log [ p e^{-4 \beta_1} + (1-p) e^{-4 \beta_2} ],
\end{equation}
as was  reported by Garrido  {\it et al}~\cite{GLM}. In two-dimensions
one has to satisfy two equations (for $\Delta E = 4$ and $8$),
\begin{eqnarray}
p e^{- 4 \beta_1} + (1-p) e^{- 4 \beta_2} & = & e^{- 4 \beta^4_{eff}}, \\
p e^{- 8 \beta_1} + (1-p) e^{- 8 \beta_2} & = & e^{- 8 \beta^8_{eff}} .
\end{eqnarray}
For fixed $p$, $\beta_1$ and $\beta_2$ each equation has the solution 
\begin{eqnarray} \label{betaeqs}
\beta^4_{eff} & = & - {1 \over 4} \log [p e^{- 4 \beta_1} + (1-p) e^{- 4 \beta_2} ] \\
\beta^8_{eff} & = & - {1 \over 8} \log [p e^{- 8 \beta_1} + (1-p) e^{- 8 \beta_2} ] 
\end{eqnarray}
which  are shown  in  Fig.~1 as  a function  of  $\beta_2$  for  fixed
$\beta_1 = 0.35$ and $p =  0.5$.  One can see  that only  for $\beta_1
=\beta_2$ is $\beta^4_{eff} = \beta^8_{eff}$. This shows that for  the
Metropolis   spin  algorithm  the  two-temperature  system  cannot  be
described by a  {\em nearest neighbor}  effective Hamiltonian.  The  same is
true for  a triangular lattice with three  (six) nearest-neighbors as
there are two (three) independent equations.

Monte Carlo  simulations show  that the configurations  generated over
time are qualitatively  similar to equilibrium configurations and  the
system exhibits an Ising-like phase transition.   We have explored the
critical behavior of this  dynamics  on  $16^2$,  $32^2$,  $64^2$  and
$128^2$ lattices.  To locate the critical point  we set $p  = 0.5$ and
$\beta_1 =  0.35$,  and searched  for the transition  as a function of
$\beta_2$.  Our best estimate  for the  critical point  is  $\beta_2 =
0.6372(5)$. The  scaling region  around this critical point appears to
be  rather  narrow,  and  the  calculation  of  critical  exponent  is
therefore quite difficult.  We choose this  particular  set of  values
for $\beta_1$ and $\beta_2$ as  they  are far from  the Ising critical
point   ($\beta_1  =  \beta_2$   =  0.440678),   and   hope  that  any
non-equilibrium effects  will be manifest.   We  computed the critical
exponents using  finite  size  scaling and Binder's cumulant analysis.
We  accumulated measurements over $5 \times 10^6$ for  $16^2$,  $10^7$
steps for $32^2$ and $64^2$ lattices and $9 \times 10^6$ for $128^2$.


The data for magnetization and  susceptibility for the four  different
lattice  sizes are given in  Table 1  and  plotted  in Figs.~2  and 3.
These figures show a change in the curvature between $\beta_2 = 0.6370$
and $\beta_2 = 0.6372$. On the basis of these data we estimate that the
critical coupling  is $\beta^*_2  = 0.6371(1)$.   The  slopes  give  an
estimate for the exponents $\beta/\nu$ and $\gamma/\nu$, assuming that
the  corrections  to the  leading  finite  size scaling forms  $m \sim
L^{-\beta/\nu}$  and $\chi  \sim L^{\gamma/\nu}$ are negligible.   The
results are
\begin{eqnarray}
\beta/\nu & = & 0.122(2) \\
\gamma/\nu & = & 1.73(2) , 
\end{eqnarray}
where the errors are determined as follows. We first compute the
statistical error in the magnetization and susceptibility for each
data point. The error in the exponents is then obtained from the mean
square fit to a straight line on a log-log plot.

The data for Binder's cumulant~\cite{BINDER},
\begin{equation}
U = 1 - {\langle M^4 \rangle \over 3 \langle M^2 \rangle^2}
\end{equation}
are also given in  Table 1 and  plotted in Figure  4 as a  function of
$\beta_2$.   The  value  of  this  cumulant  at the  critical point is
conjectured  to   be  a   universal  number   independent  of  lattice
size~\cite{BINDER}.  Figures 5a and b show the cumulant values for the
two pairs of lattice sizes, ($32^2$, $64^2$) and ($64^2$, $128^2$), at
different temperatures around the critical point.   The solid straight
is  the best fit to the  data  and  the  dashed  line  in the  figures
corresponds to $U_{L/2} =  U_{L}$.  The point of crossing of these two
lines  gives  an  estimate of the critical point.  The  crossing  takes
place at
\begin{eqnarray}
U^*({\rm 32\ vs\ 64})  & = &  0.610(2)   \\
U^*({\rm 64\ vs\ 128}) & = &  0.605(10)   
\end{eqnarray}
and the  corresponding  estimates for $\beta_2^*$  are $0.6370(2)$ and
$0.6360(12)$  respectively.   These  values  are  consistent  with the
estimate from finite  size scaling given above.  Our longest runs were
done at $\beta_2 = 0.6372$, at which temperature our estimates
\begin{eqnarray}
U_{16}  & = &  0.611(1)   \\
U_{32}  & = &  0.611(2)   \\
U_{64}  & = &  0.612(6)   \\
U_{128} & = &  0.615(7)
\end{eqnarray}
compare  very  well with the value  $U^* = 0.611(1)$ computed by Bruce
~\cite{BRUCE}  for  the $2D$ equilibrium Ising  model, and $U^*_{64} =
0.611(5)$  independently  by  us.   Thus,  we  shall  henceforth  call
$\beta_2 = 0.6372$ the critical point.

>From the scaling of the cumulant one can obtain an estimate of $\nu$,
\begin{equation}
\left. {d U_L \over d \beta } \right|_{\beta_c} = L^{1/\nu} G( L^{1/\nu} \epsilon)
\end{equation}
where $\epsilon  = (\beta_c  -  \beta)/\beta$.   To  do  this we first
compute the slope $d U_L/d \beta$  for $L=32,\ 64$ and $128$ using the
data  points near the critical point  and then fit these values versus
$L$ using the above expression to obtain an estimate for $\nu$.  We find,
\begin{equation}
\nu = 0.99(5) .
\end{equation}
Another estimate can be computed from $d U_L/ d U_{L'}$ in the
critical region because this quantity should scale as 
\begin{equation}
{ d U_L \over d U_{L'}} \sim \left ({L \over L'} \right)^{1/\nu}  .
\end{equation}
Using the more precise data  shown in Fig.~5a for  $L=32$  and $64$ we
obtain $\nu=0.95(8)$.   The  agreement between these  results  and the
values  for  the  equilibrium  Ising  model  is quite  good  ($\nu=1$,
$\beta/\nu=0.125$,  $\gamma/\nu=1.75$,  $U^*=0.611(1)$)  and  provides
strong evidence for their equivalence.

To further  confirm  this  equivalence  we  measured  the  probability
distribution  for the  magnetization and energy at $\beta_1=0.35$  and
$\beta_2=0.6372$ and compare them with those for the equilibrium model
in  Figs.~7a  and  b.   The  agreement  is  somewhat  better  for  the
magnetization than for the energy and qualitatively  the  distribution
functions are equilibrium-like.  The data are slightly more disordered
than  those  for  the  critical  equilibrium  model,  suggesting  that
$\beta_2^*$ may be slightly larger than $0.6372$.

The bond version of the  Metropolis dynamics  has not been studied  as
extensively, and  we postpone its discussion until Section  5 where we
compare the different dynamics.

\bigskip
\bigskip
{\Large \bf 4. Swendsen-Wang non-equilibrium dynamics}
\bigskip

We start with the bond contribution to a global probability distribution,
\begin{equation}
P_{ij}  = p e^{-\beta_1 (1 - \sigma_i \sigma_j)} + (1-p) e^{-\beta_2 (1 - \sigma_i \sigma_j)}
\end{equation}
If we assume that $P_{ij}$ is always equivalent to an equilibrium
distribution with coupling $\beta_{\rm eff}$, $i.e.$ 
\begin{equation} \label{prob}
P_{ij}  = p e^{-\beta_1 (1 - \sigma_i \sigma_j)} + (1-p) e^{-\beta_2 (1 - \sigma_i \sigma_j)} = e^{- \beta_{\rm eff} (1 - \sigma_i \sigma_j)} , 
\end{equation}
then there exists a solution satisfying this equation for the two 
possible values of the bond energy, 
\begin{eqnarray}
p e^{- 2 \beta_1} + (1 - p) e^{- 2 \beta_2} & = & e^{- 2 \beta_{\rm eff}},
\;\;\;\;\;\; [\sigma_i = - \sigma_j]\\
p  + (1 - p)  & = & 1,  \;\;\;\;\;\; [\sigma_i = \sigma_j] .
\end{eqnarray}
The solution is,
\begin{equation}
 p = {e^{-2 \beta_{\rm eff}} - e^{-2 \beta_2} \over e^{-2
\beta_1} - e^{-2 \beta_2}},
\end{equation}
or  equivalently,  for  a set of values  $\beta_1$, $\beta_2$ and $p$,
there always exists a $\beta_{\rm eff}$ given by,
\begin{equation} \label{spider}
\beta_{\rm eff} = - {1 \over 2} \log [ p (e^{-2 \beta_1} - e^{-2 \beta_2})
+ e^{-2 \beta_2} ]
\end{equation}
that corresponds  to  an  Ising equilibrium system. This implies  that
this  dynamics  is  nothing  more  than  equilibrium  dynamics  in
disguise.  If we set $\beta_{\rm  eff}  = \beta_c$  we  find  lines of
critical points in the $\beta_1 - \beta_2$ plane given by
\begin{equation} \label{spider2}
\beta_1 = - {1 \over 2} \log [ {1 \over p}( e^{-2 \beta_c} - (1 - p)
e^{- 2 \beta_2} ) ] .
\end{equation}
A set of  these critical lines is shown in Fig.~6 for different values
of $p$. Our simulations confirm this equivalence. Notice that for some
extreme values of  $\beta_1$ ($\beta_2$) there is no positive value of
$\beta_2$ ($\beta_1$) that yields critical  behavior. 

The reason for the  equivalence  is  that for  the Swendsen-Wang  bond
dynamics there are, independent of  the number of spatial  dimensions,
only two equations  of constraint as each  bond  is  independently  in
contact with the heat bath.  One is the trivial condition $p + (1-p) =
1$ and the second is the desired result given in Eq.~\ref{spider}.  In
a different  context,  a similar analysis of different ways to satisfy
local equations for block  percolation in equilibrium systems has been
made in ref~\cite{BT}.

Lastly, we have  measured  autocorrelation times  at criticality  with
$\beta_1 =  0.1$,  $\beta_2 = 2.318$, and $p=0.5$  ($\beta_{\rm eff} =
0.440687$) to see if the stochastic choice of temperatures accelerates
the decorrelation process. We find that the autocorrelation times  are
comparable to the  values  for  the standard equilibrium Swendsen-Wang
values.  This is as expected since the two models are locally equivalent. 
 
\bigskip
\bigskip
{\Large \bf 5. Other dynamics and comparisons}
\bigskip

We have investigated the spin and bond version of Glauber dynamics for
only  three combinations  of  $\beta_1$  and  $\beta_2$  using  $32^2$
lattices.   The  ensemble size in  these  runs is $\sim 50,000$ update
sweeps,  significantly  smaller than  for  Metropolis or Swendsen-Wang
dynamics.  In Figs.~8a, 8b, and 8c we compare  the equation  of  state
($\langle  E\rangle$  {\it vs} $\langle m^2 \rangle$)  for the various
dynamics  for  the  three  different  combinations  of  $\beta_1$  and
$\beta_2$.  Figure 8a shows the results for $\beta_1  = 0.4$, $\beta_2
= 0.6$,  Figure 8b corresponds to  $\beta_1  = 0.1$, $\beta_2 = 2.318$
and Figure 8c to  $\beta_1 = 0.2$, $\beta_2  = 0.738464$.  These three
sets  of  temperatures   were  chosen   such  that  the  corresponding
$\beta_{\rm  eff}$  for  Swendsen-Wang  bond dynamics corresponds to a
cold ($\beta_{eff} = 0.49$) , critical ($\beta_{eff} = 0.4406868$) and
hot ($\beta_{eff} = 0.40$) system  respectively.  All  the simulations
were done with $p=0.5$.  The dashed line corresponds to the result for
the equilibrium Ising model obtained on $L=32^2$ lattices.  The errors
in the data are roughly equal to the size of the symbols.

>From Figs.~8a-c it is clear that  the equation of state  depends  very
sensitively on the dynamics and there is a large spread in the results
for  all three choices of  temperatures.  In all cases, except for the
Swendsen-Wang spin version,  the results lie  very  close  to the line
characterizing  the equilibrium Ising model.  The deviations from  the
equilibrium results  in all  three versions of the  2-temperature spin
dynamics are in the direction of a more ordered system.  The situation
is reversed for  the bond dynamics; compared to the equilibrium values
the two-temperature results are less ordered.  Qualitatively, the data
with  the  various  dynamics  show the  following ordering: Metropolis
spin,  Glauber  spin, Metropolis bond, Glauber bond, and Swendsen-Wang
bond.  This pattern is also shown  in Figs.~9a and b where we give the
magnetization and  energy probability distributions for $\beta_1=0.35$
and  $\beta_2=0.6372$  (a  critical  point  for  the  Metropolis  spin
dynamics).  All cases  appear  to have the  same  functional form  but
their   position  and  amplitudes  are  rescaled.   The  magnetization
probability distributions  are Gaussian  near the peak  but have  long
tails toward $m=0$.  Further work is needed to explore the possibility
that all of these probability distributions are just rescaled forms of
the equilibrium  distribution and correspond  to different $\beta_{\rm
eff}$'s

\bigskip
\bigskip
{\Large \bf 6. Three-Temperature Model}
\bigskip

To study  the general case of the competition of many  temperatures we
extended  the  analysis  to   the  three-temperature  Metropolis  spin
dynamics.  The three temperatures are chosen with probabilities $p_1$,
$p_2$ and $p_3$:
\begin{equation}
W(\sigma' \leftarrow \sigma) =  p_1 W_1(\sigma' \leftarrow \sigma)
 + p_2 W_2(\sigma' \leftarrow \sigma)  + p_3 W_3(\sigma' \leftarrow \sigma)  .
\end{equation}
We  fixed  $\beta_1=0.35$, $\beta_3=0.6372$, and $p_1=p_2=p_3=1/3$ and
varied  $\beta_2$  about  the   equilibrium  critical  value  $\beta_2
=0.4406868$.  We expected  the system to display critical behavior for
$\beta_2 =0.4406868$.

The system effectively displays  critical behavior in  that  region as
can be seen  in Fig.~10 where results for  Binder's cumulant are shown
for  $\beta_2$  in  the  range  $[0.424,\  0.456$.   Furthermore,  the
probability distribution functions for $\beta_2=0.4406868$  match  the
ones for the  two-temperature  model  discussed  in Section 4  and the
equilibrium model  at  criticality as can be seen in Figs.~11a and b.
We  do find an apparent narrowing of the critical region compared with
the two-temperature models and the statistical quality of the data are
not accurate enough to measure the exponents.

Based on the study  of the 3  temperature model we make the  following
conjecture  for the Metropolis spin  dynamics. A model in contact with
an arbitrary number of heat-baths will display Ising critical behavior
provided each  temperature or  a pair of them is tuned to the critical
value.  As the  number  of pairs of temperatures increase the critical
region becomes narrower,  making the measurement of critical exponents
and the study of critical behavior more difficult.

\bigskip
\bigskip
{\Large \bf 7. Conclusions}
\bigskip

We present high  statistics  results showing  that  for the Metropolis
spin dynamics the stationary  states produced  by the  two-temperature
model  are  very  similar to  equilibrium  states.   On basis  of  the
agreement  of the critical exponents and Binder's cumulant we conclude
that  the  two-temperature  Metropolis  spin  dynamics is in the  same
universality  class as the  Ising model.  We also  show that  the bond
version of  the  two-temperature Swendsen-Wang dynamics  can be mapped
into  an  equilibrium  Ising  model   at  an   intermediate  effective
temperature.   Thus,  for these  cases  our  results  agree  with  the
conjecture  of   Grinstein  {\it  et  al.}  that  any  non-equilibrium
spin-flip dynamics which  preserves up-down symmetry  belongs  to  the
same universality class as the equilibrium Ising model.  Assuming that
the  system  evolves  into  an  equilibrium  distribution  after  some
thermalization steps, one can measure the critical  exponents and flow
of renormalized  couplings using the Monte Carlo renormalization group
method.  We hope to investigate this possibility in the future.

Our  results for  the  Metropolis  bond,  Glauber spin and  bond,  and
Swendsen-Wang spin dynamics are qualitative.  Further work is required
to confirm that they too belong to the same  universality class as the
equilibrium Ising model.

We have extended the two-temperature critical Metropolis spin dynamics
to the three temperature case.  We find that the system shows critical
behavior  when the third temperature is  tuned  to  $\beta=0.4406868$.
Based on  this we  conjecture  that the  Ising  critical  behavior  is
preserved  as  long as  one adds  pairs of temperatures  that are,  by
themselves, critical.  The critical  region appears to become narrower
as the  number  of pairs  of temperature values are  increased and the
statistical  quality   of  the  data  deteriorates.   This  makes  the
calculation of critical  exponents and  the study  of the models  more
difficult.



\bigskip
\bigskip
{\Large \bf Acknowledgments}
\bigskip

We thank P.~L.~Garrido for helpful comments and
M.~A.~Mu\~noz-Martinez, R.~Mainieri, R.~Swendsen, X.~Wang, T,
Bhattacharya, G.  Grandy and N.~Kawashima for discussions and
B.~J.~Alder, S.~Chen, G.~D.~Doolen and J.~Mesirov for support.  We
also wish to acknowledge the Advanced Computing Laboratory of Los
Alamos National Laboratory and Thinking Machines Corp. for support of
the computations performed here.

\newpage

\newpage

Table I: Data for Susceptibility, Magnetization and Binder's cumulant 
for the 2 temperature Metropolis spin dynamics ($\beta_1=0.35$, $p=0.5$)

\begin{center}
{\small
\begin{tabular}{|l|l|l|l|l|} \hline
$\beta_2$ & $L$  &    $\chi$   &   $|m|$    &   $U$     \\ \hline
0.6372  &   16  &     9.11(5)  &  0.7043(3) &    0.611(1) \\ \hline
0.6300  &   32  &     33.5(3)  &  0.629(2)  &    0.603(3)   \\
0.6350  &   32  &     32.0(4)  &  0.639(2)  &    0.608(4)    \\
0.6370  &   32  &     30.2(8)  &  0.647(3)  &    0.611(11)   \\
0.6372  &   32  &     30.6(3)  &  0.6464(7) &    0.611(2)   \\
0.6375  &   32  &     30.6(3)  &  0.6463(8) &    0.611(3)   \\
0.6400  &   32  &     30.0(6)  &  0.651(2)  &    0.613(7)   \\
0.6475  &   32  &     27.2(4)  &  0.666(1)  &    0.619(3)   \\
0.6500  &   32  &     26.3(6)  &  0.671(2)  &    0.621(5)   \\
0.6550  &   32  &     25.1(4)  &  0.679(1)  &    0.624(4)   \\
0.6600  &   32  &     23.3(7)  &  0.689(2)  &    0.628(5)    \\ \hline 
0.6300  &   64  &     114(4)   &  0.566(4)  &     0.600(13)    \\
0.6350  &   64  &     112(5)   &  0.580(4)  &     0.605(12)   \\
0.6370  &   64  &     103(2)   &  0.592(2)  &     0.611(4)   \\
0.6372  &   64  &     101(2)   &  0.594(2)  &     0.612(6)   \\
0.6375  &   64  &     101(2)   &  0.594(2)  &     0.612(5)   \\
0.6400  &   64  &     96(5)    &  0.605(2)  &     0.616(3)   \\
0.6475  &   64  &     76(5)    &  0.635(4)  &     0.629(14) \\
0.6500  &   64  &     68(3)    &  0.644(3)  &     0.633(11)    \\
0.6550  &   64  &     64(5)    &  0.655(3)  &     0.636(11)    \\
0.6600  &   64  &     56(3)    &  0.669(2)  &     0.640(4)    \\ \hline 
0.6350  &   128 &     409(51)    &  0.52(1)   &      0.597(40)   \\
0.6370  &   128 &     357(22)    &  0.538(6)  &      0.607(22)   \\
0.6372  &   128 &     315(12)    &  0.550(2)  &      0.615(7)  \\
0.6375  &   128 &     328(18)    &  0.550(3)  &      0.615(11)   \\
0.6400  &   128 &     257(20)    &  0.573(5)  &      0.627(21)   \\
0.6475  &   128 &     197(19)    &  0.611(3)  &      0.639(9)  \\
0.6500  &   128 &     148(10)    &  0.626(3)  &      0.645(10)   \\
0.6550  &   128 &     133(15)    &  0.644(3)  &      0.649(9)  \\
0.6600  &   128 &     91(10)     &  0.664(2)  &      0.654(7)  \\  \hline 
\end{tabular}
}
\end{center}

\newpage

\overfullrule=0pt
\def\hsbody{391.46176pt}
\long\def\pageinsert#1\endinsert{#1}

\ifx\ifstretchtofull\relax\message{Attempt to redefine eps macros ignored}%
\else
 \input epsf
 \newif\ifstretchtofull{\toks0={\fi}}%
 \def\epsfigure#1#2{
   \pageinsert\parindent=0pt 
   $$\epsfxsize=\hsize\ifstretchtofull\epsfysize=\hsbody\fi\epsfbox{#1}$$
   \par\bigskip\bigskip\bigskip #2 \vfill\endinsert}
\fi

\stretchtofulltrue 

\stretchtofullfalse

 \epsfxsize=330pt \epsfxsize=330pt \epsfbox{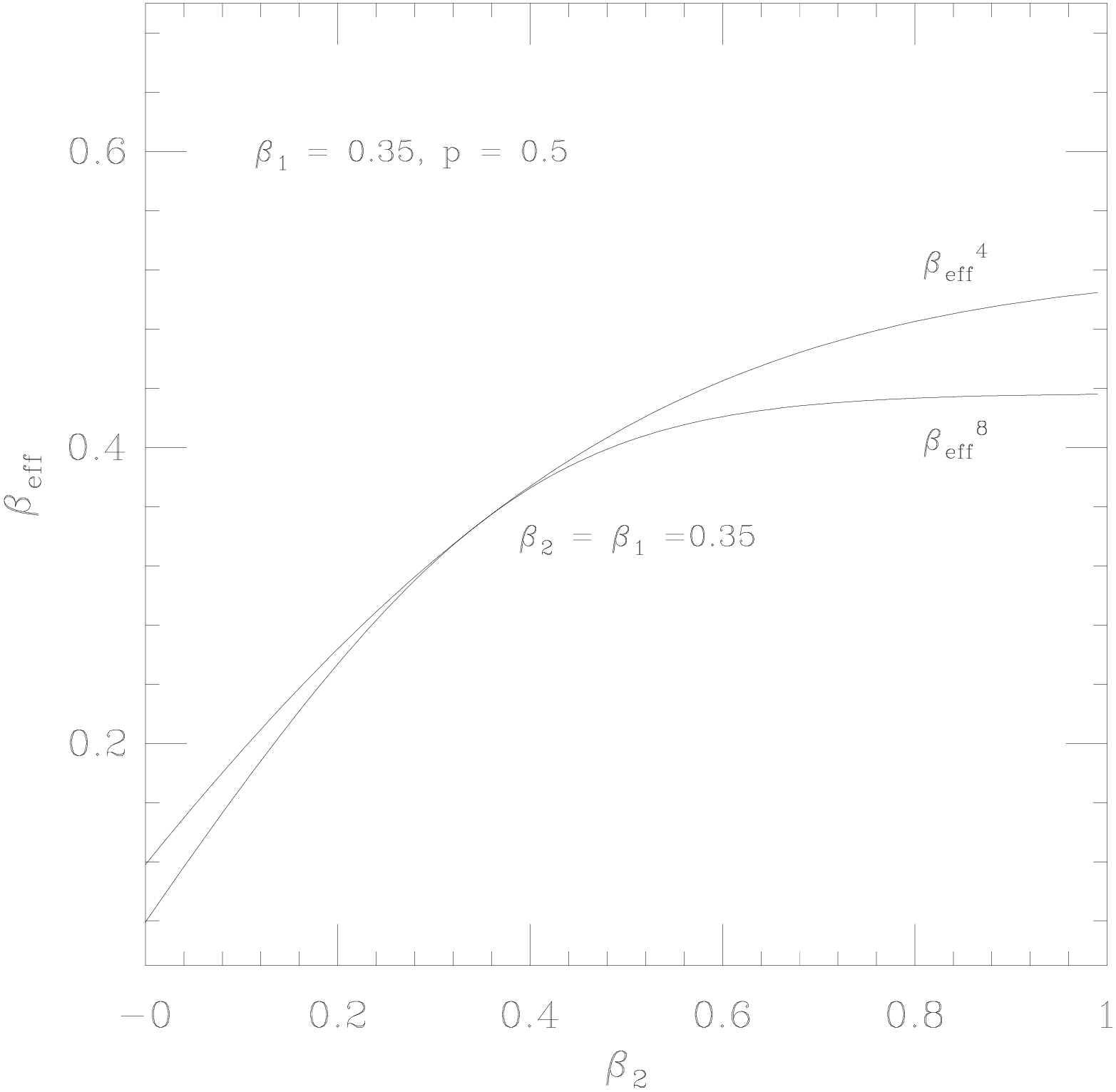}

{\bf  Fig. 1.-} Plot of $\beta^4_{eff}$  and
$\beta^8_{eff}$ (see eq.~\ref{betaeqs}) versus $\beta_2$  (at $\beta_1
=  0.35$  and $p  =  0.5$)  for the  non-equilibrium  Metropolis  spin
dynamics.

 \epsfxsize=330pt \epsfxsize=330pt \epsfbox{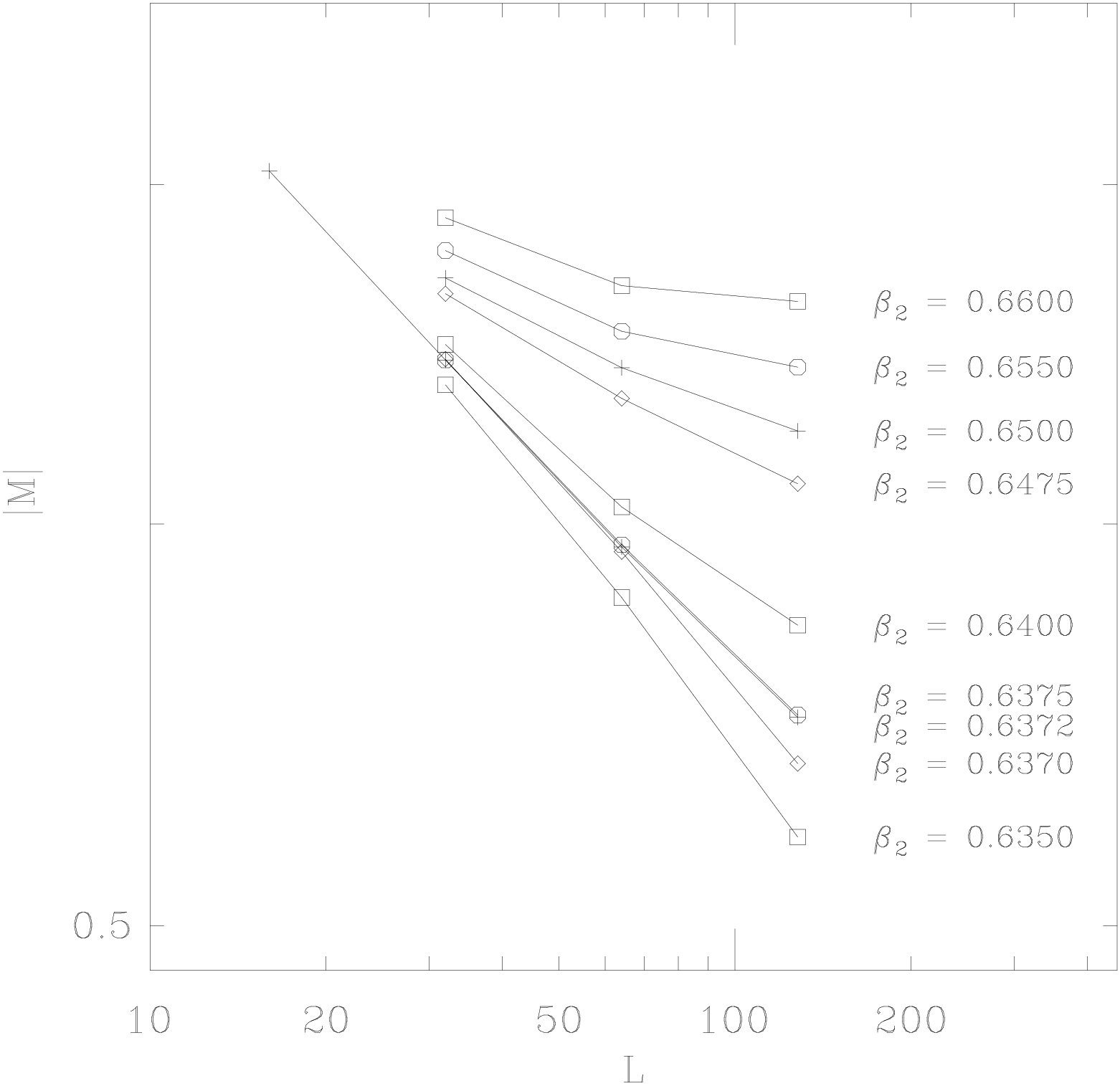}
{\bf Fig. 2.-} Scaling behavior of  Magnetization
as  a function  of  $\beta_2$ (with $\beta_1 = 0.35$ and  $p=0.5$ held
fixed) on different lattice sizes  for the  non-equilibrium Metropolis
spin dynamics.

 \epsfxsize=330pt \epsfxsize=330pt \epsfbox{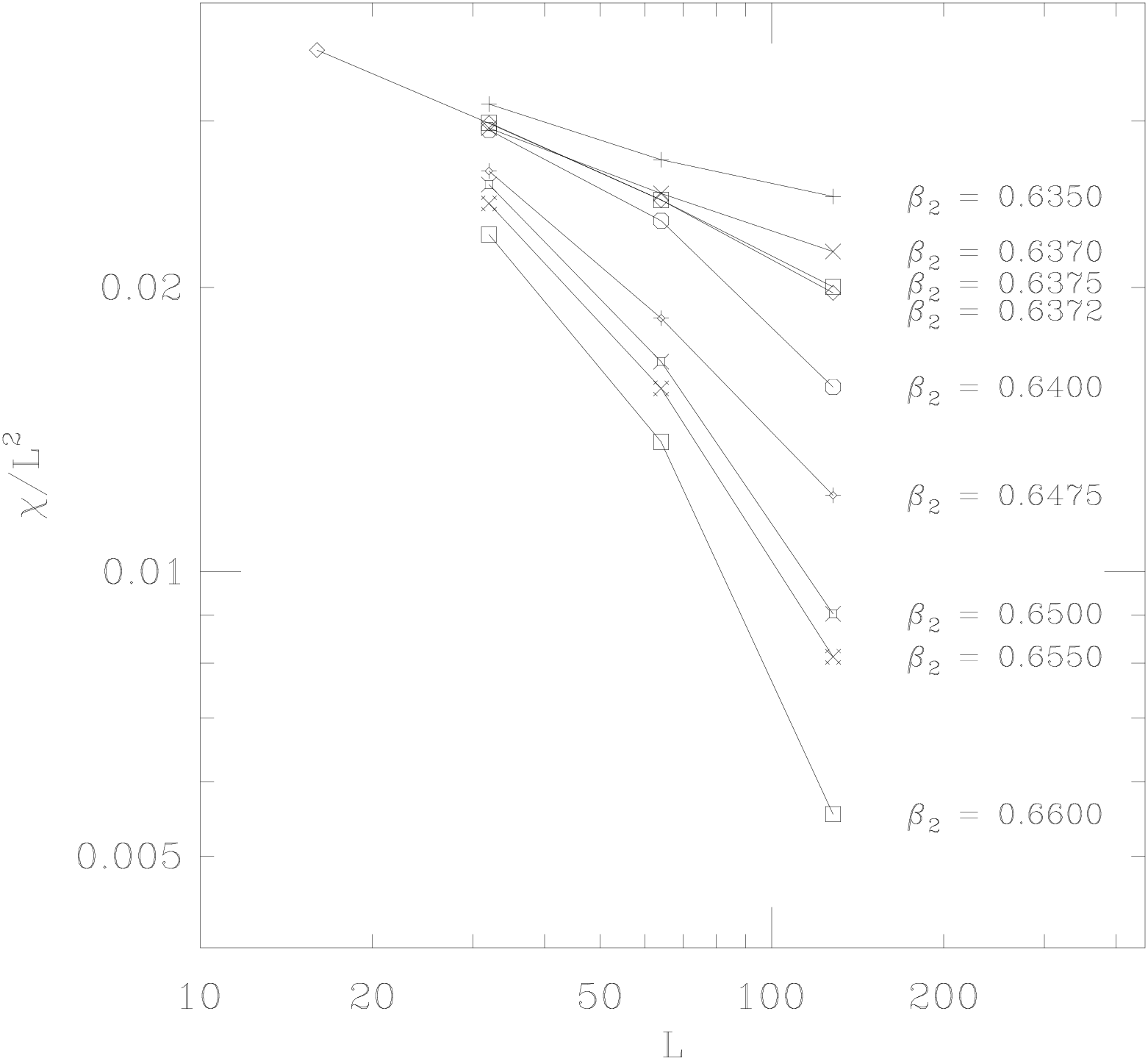}
{\bf   Fig.   3.-}   Scaling   behavior   of
susceptibility  as a function of $\beta_2$ (with  $\beta_1 = 0.35$ and
$p=0.5$   held  fixed)   on    different   lattice   sizes   for   the
non-equilibrium Metropolis spin dynamics.

 \epsfxsize=330pt \epsfxsize=330pt \epsfbox{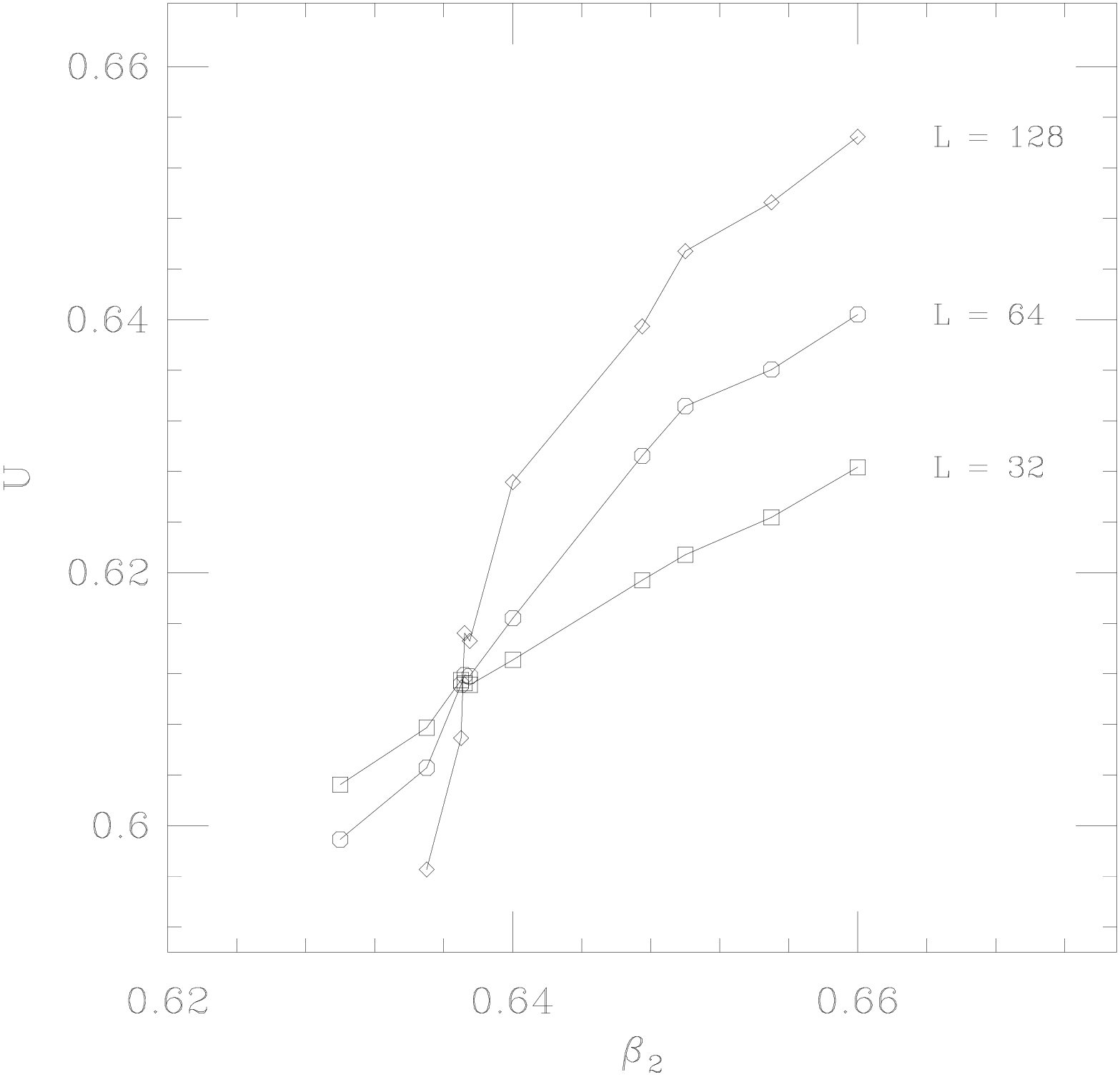}
{\bf Fig. 4.-} Cumulant values as a function of
$\beta_2$  (at  $\beta_1 = 0.35$ and $p=0.5$) and on different lattice
sizes for the non-equilibrium Metropolis spin dynamics.

\epsfigure{ 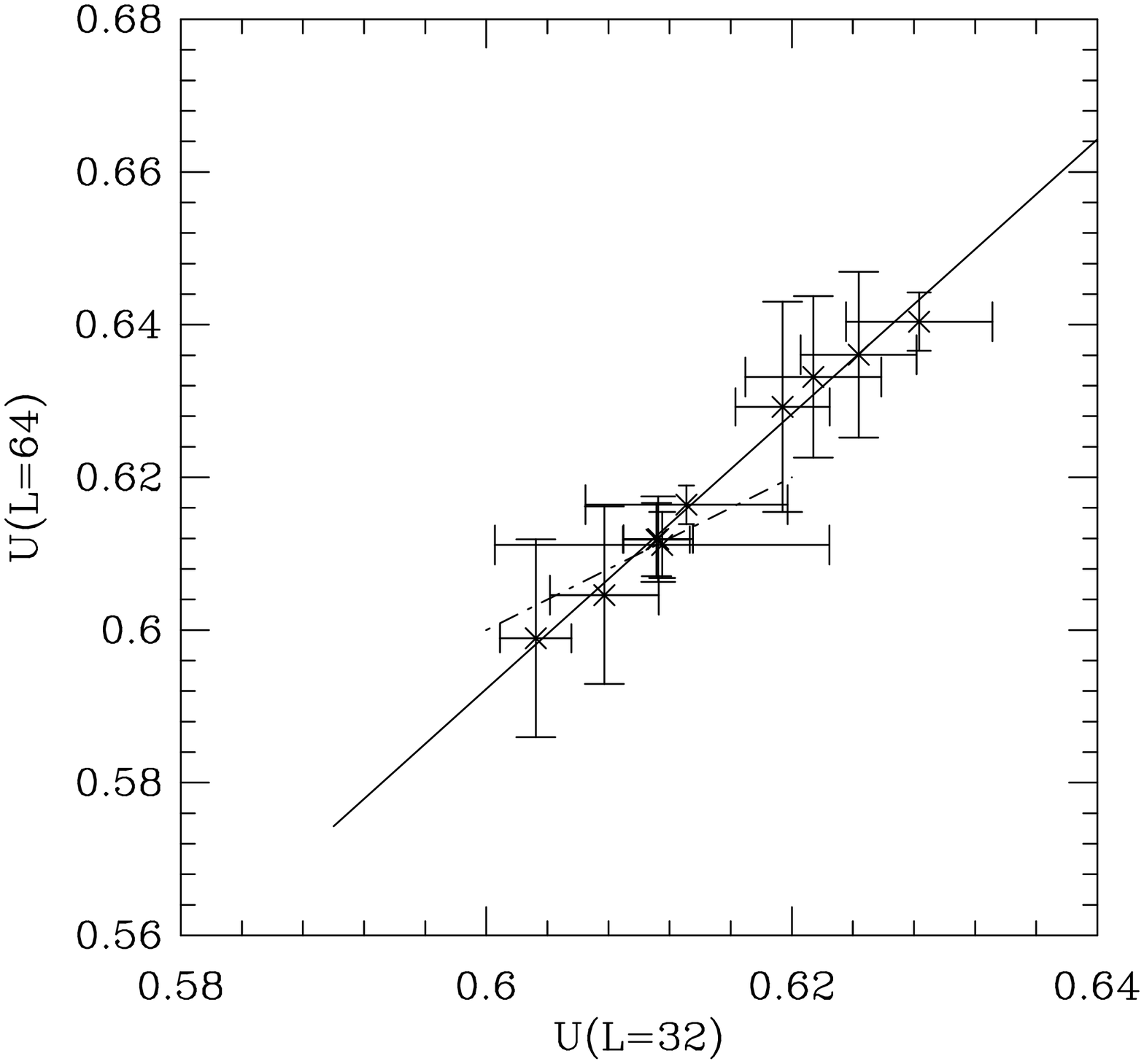}{ {\bf Fig.  5a.-} Cumulant values for $32^2$ versus
$64^2$  lattices  at  temperatures  around the critical point for  the
non-equilibrium Metropolis spin dynamics.}

\epsfigure{ 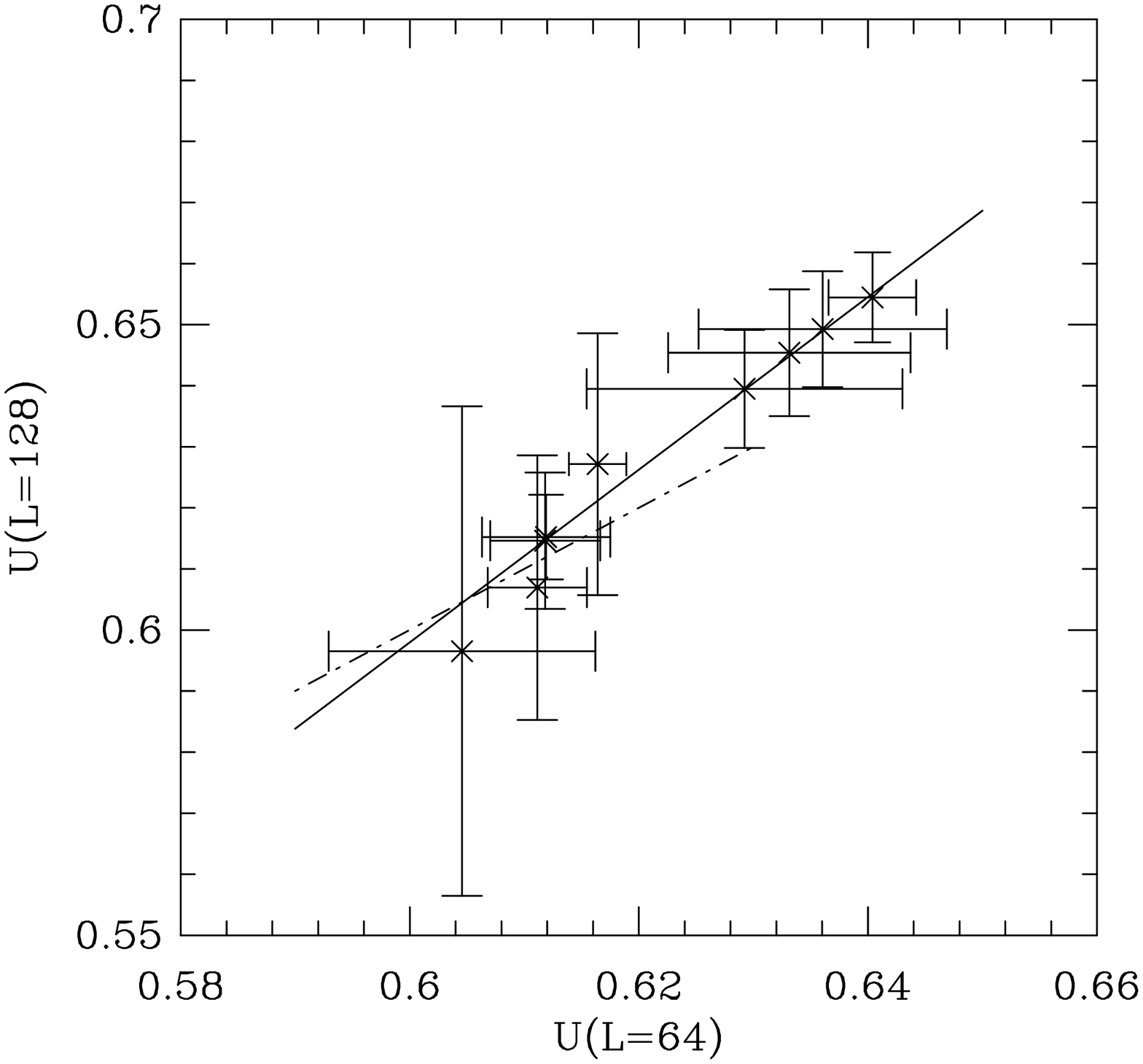}{ {\bf Fig. 5b.-} Cumulant values for $64^2$ versus
$128^2$  lattices  at temperatures around the critical  point for  the
non-equilibrium Metropolis spin dynamics.}


 \epsfxsize=330pt \epsfxsize=330pt \epsfbox{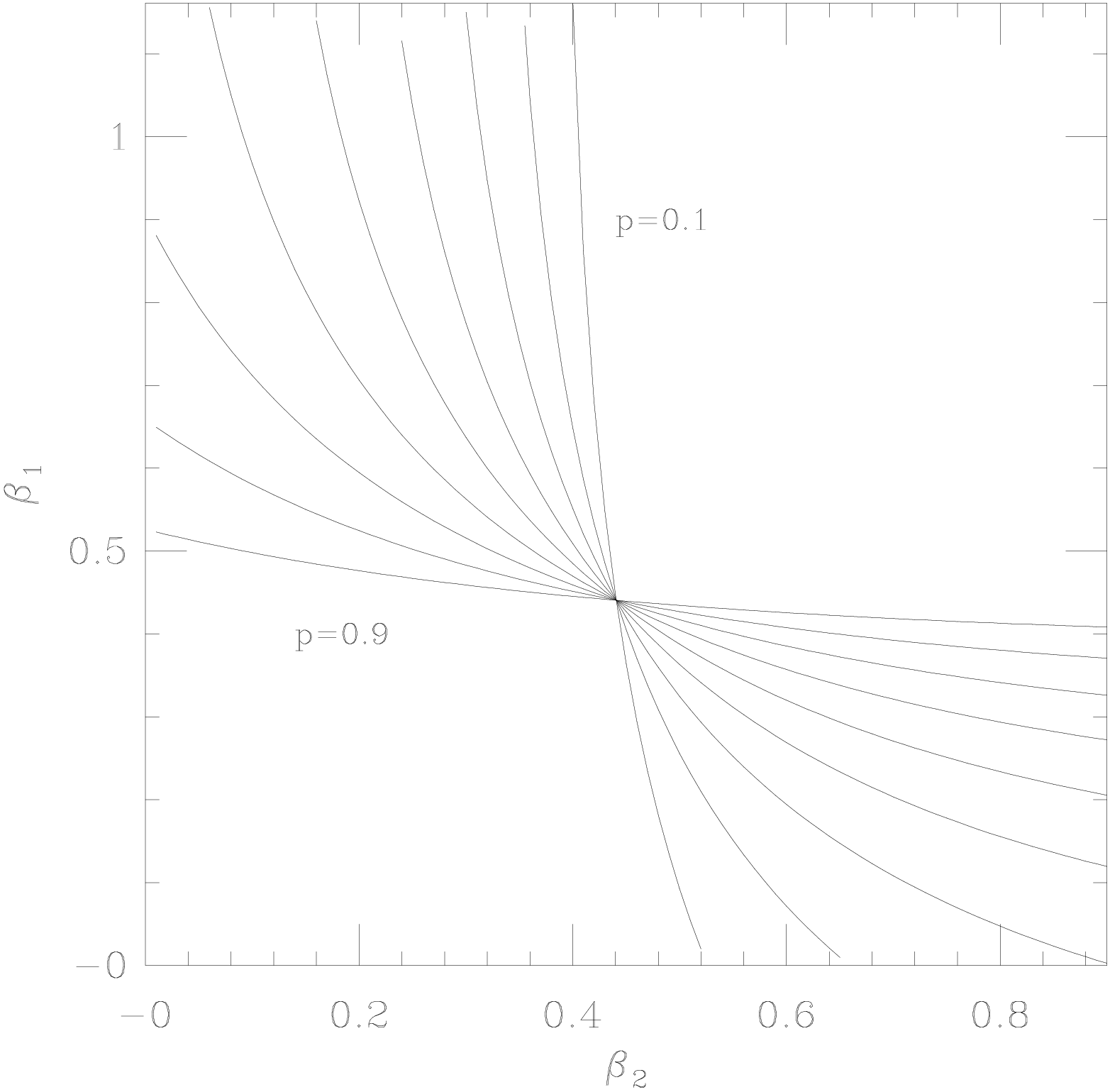}
{\bf Fig. 6.-}  Critical   lines   (see
eq.~\ref{spider2}) for the Swendsen-Wang bond dynamics.

 \epsfxsize=330pt \epsfxsize=330pt \epsfbox{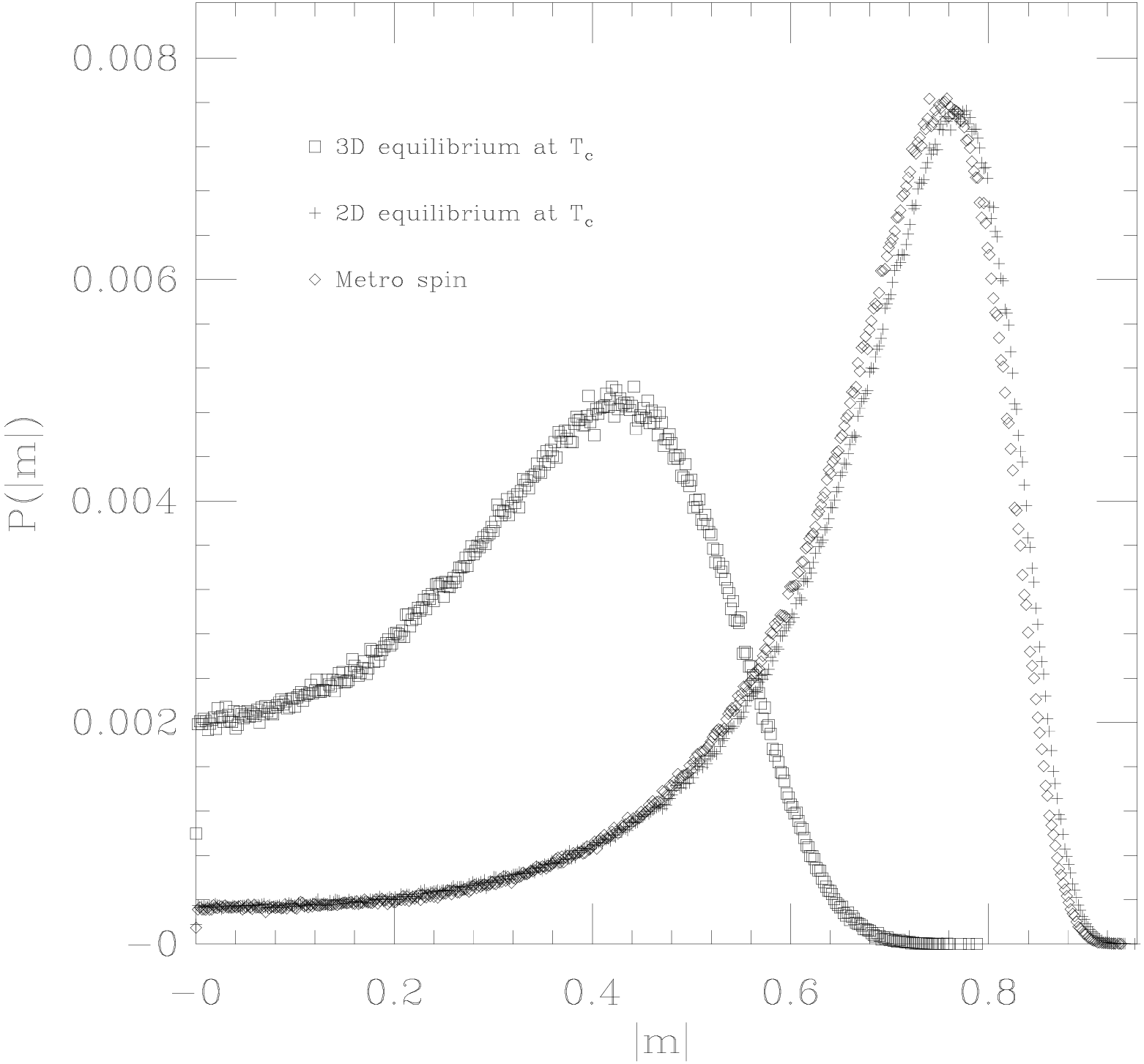}
{\bf Fig. 7a.-} Magnetization probability
distributions $P(|m|)$ for the Metropolis spin dynamics at $\beta_1 =
0.35$, $\beta_2 = 0.6372$, $p=0.5$ and $L=32$ (critical point).  The
probability distributions for the $2D$ and $3D$ equilibrium Ising
model are also shown for comparison.

 \epsfxsize=330pt \epsfxsize=330pt \epsfbox{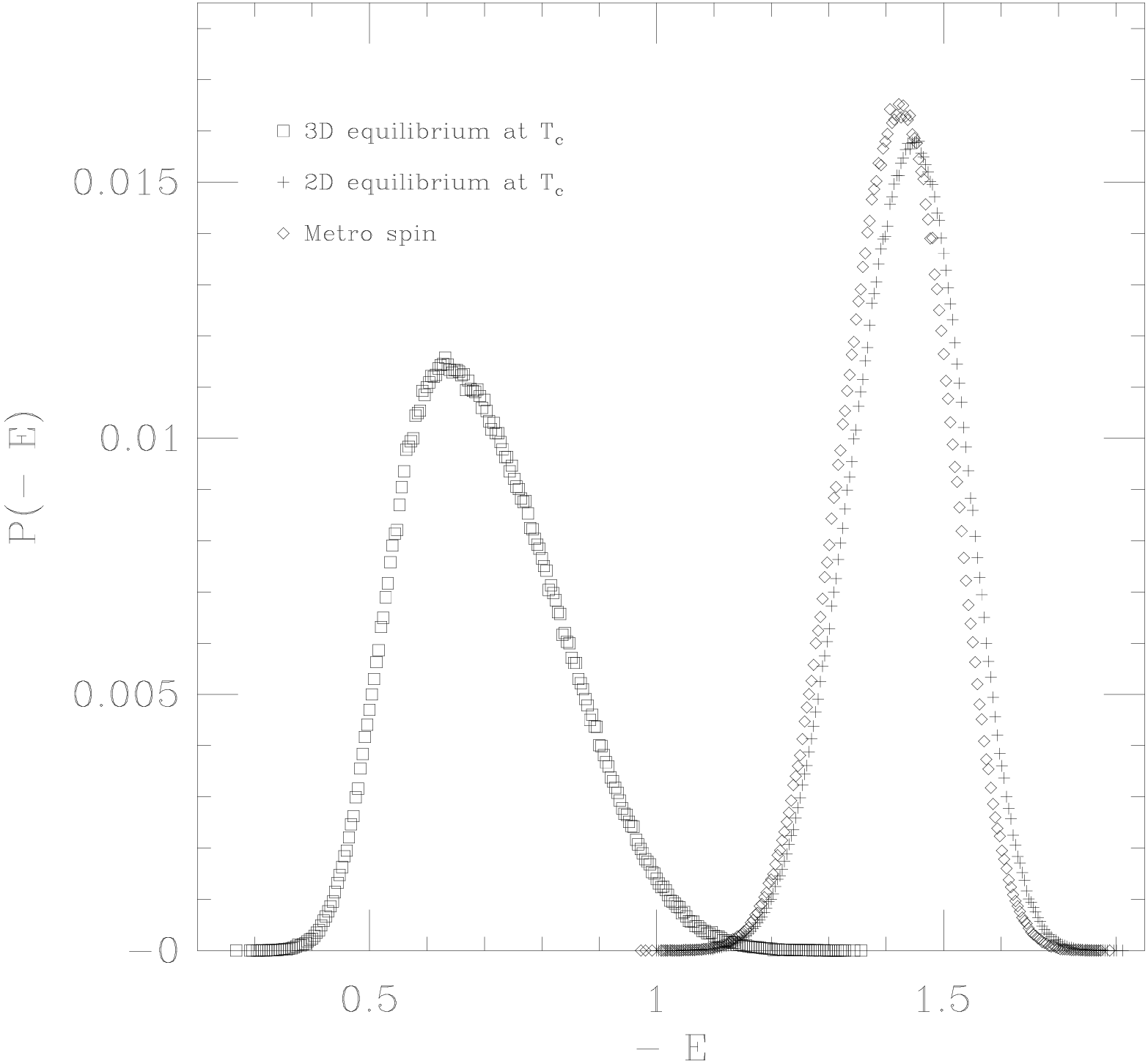}
{\bf Fig. 7b.-} Energy probability
distributions $P(-E)$ for the Metropolis spin dynamics at $\beta_1 =
0.35$, $\beta_2 = 0.6372$, $p=0.5$ and $L=32$ (critical point).  The
probability distributions for the $2D$ and $3D$ equilibrium Ising model at
criticality are also shown for comparison.

\stretchtofulltrue

\epsfigure{ 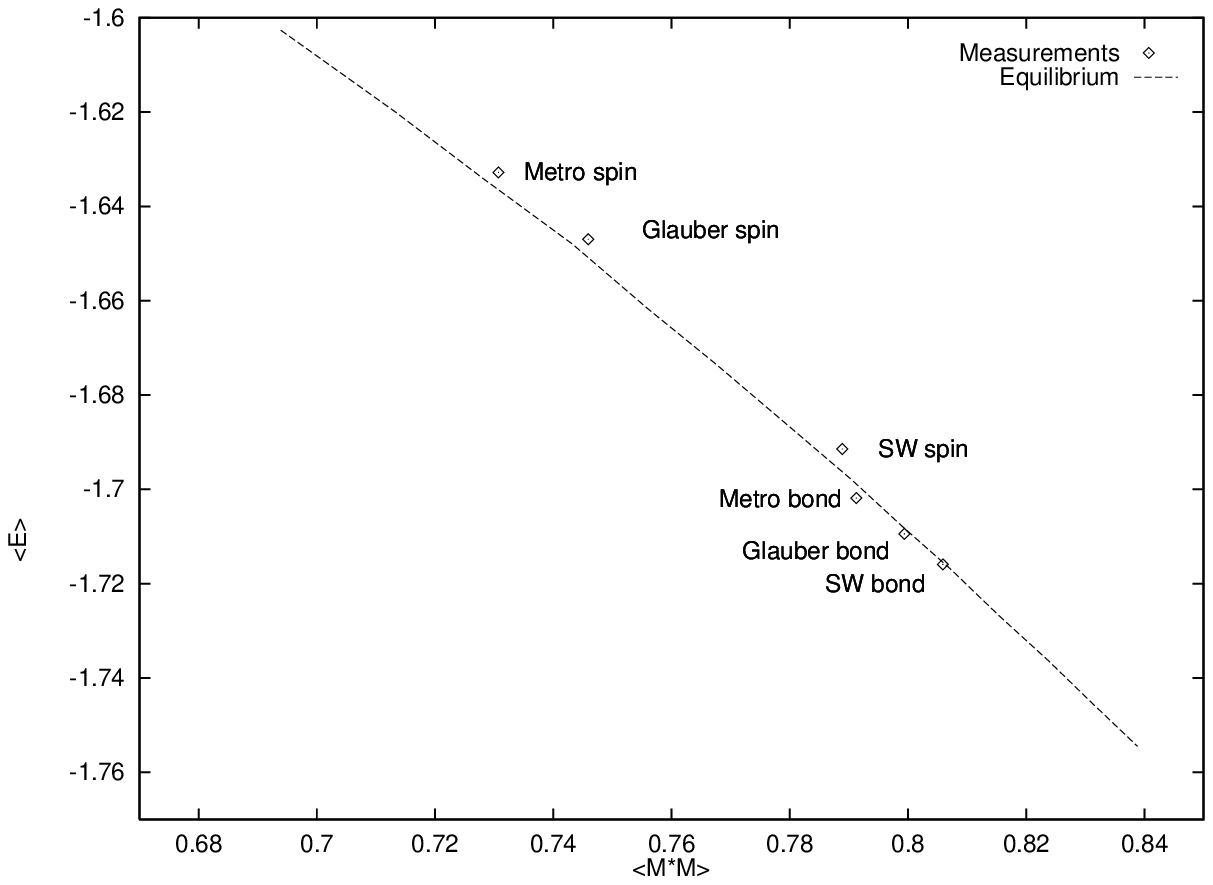}{ {\bf Fig.   8a.-}  Equation  of  state  data  for
different dynamics with $\beta_1 = 0.4$, $\beta_2 = 0.6$, $p=0.5$  and
$L=32$ ($\beta_{\rm eff}^{SW_{bond}} = 0.49$)}

\epsfigure{  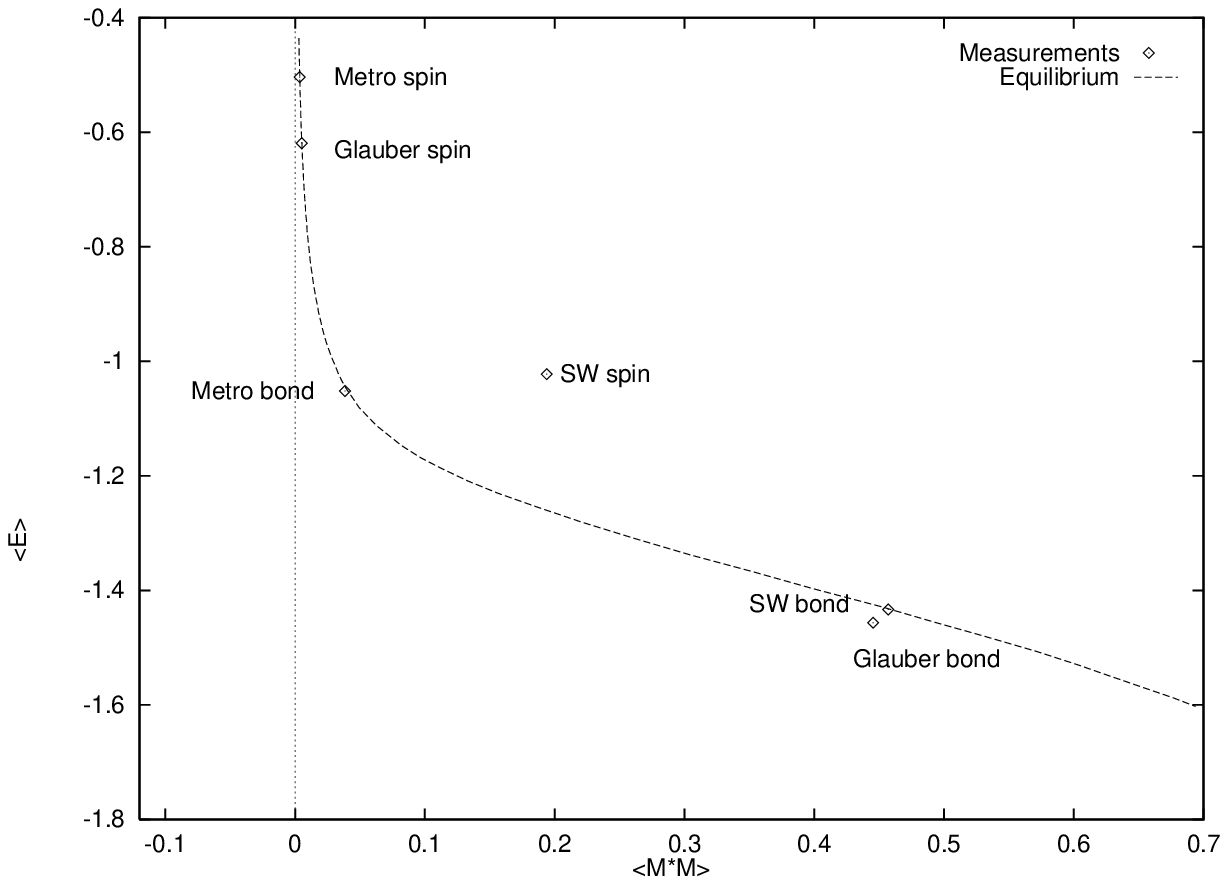}{ {\bf  Fig.  8b.-} Equation  of  state  data  for
different  dynamics with $\beta_1 = 0.1$, $\beta_2 =  2.318$,  $p=0.5$
and $L=32$ ($\beta_{\rm eff}^{SW_{bond}} = 0.4406868$)}

\epsfigure{  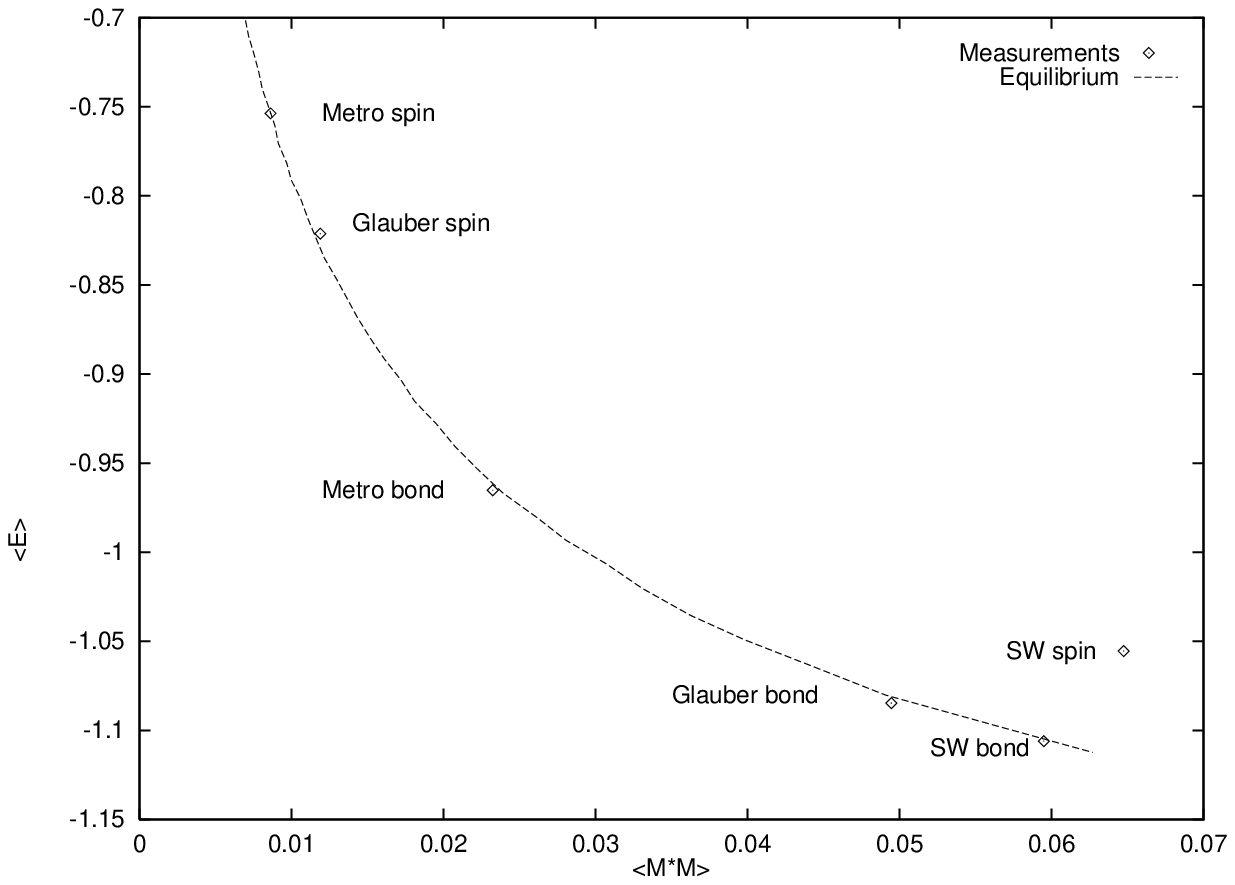}{ {\bf  Fig.  8c.-}  Equation  of  state  data  for
different dynamics with $\beta_1 = 0.2$, $\beta_2 = 0.738464$, $p=0.5$
and $L=32$ ($\beta_{\rm eff}^{SW_{bond}} = 0.4$)}

\stretchtofullfalse

 \epsfxsize=330pt \epsfxsize=330pt \epsfbox{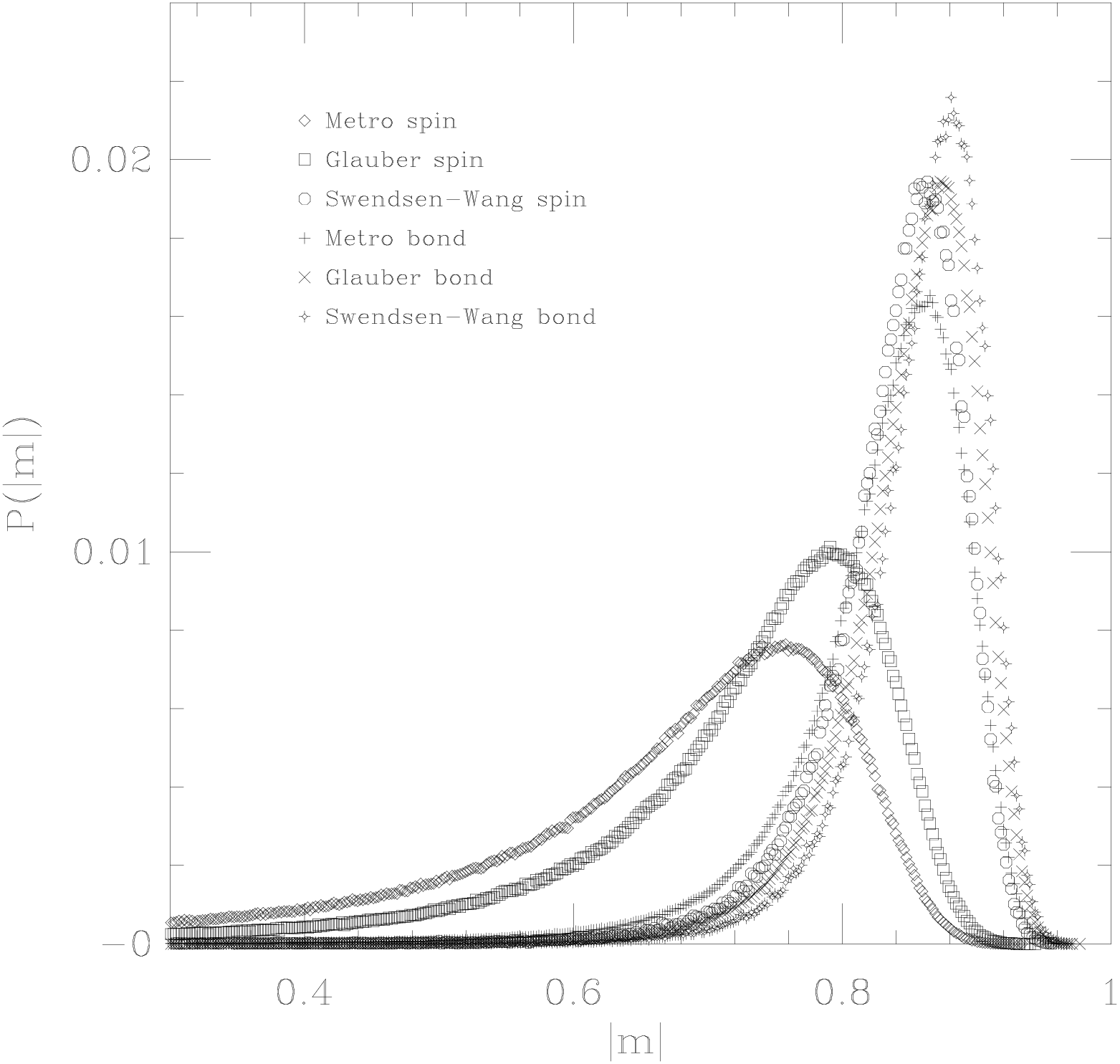}
{\bf Fig. 9a.-} Magnetization probability
distributions $P(|m|)$ for the spin and bond versions of Metropolis,
Glauber and Swendsen-Wang dynamics at $\beta_1 = 0.35$, $\beta_2
= 0.6372$, $p=0.5$ and $L=32$ (critical point).

 \epsfxsize=330pt \epsfxsize=330pt \epsfbox{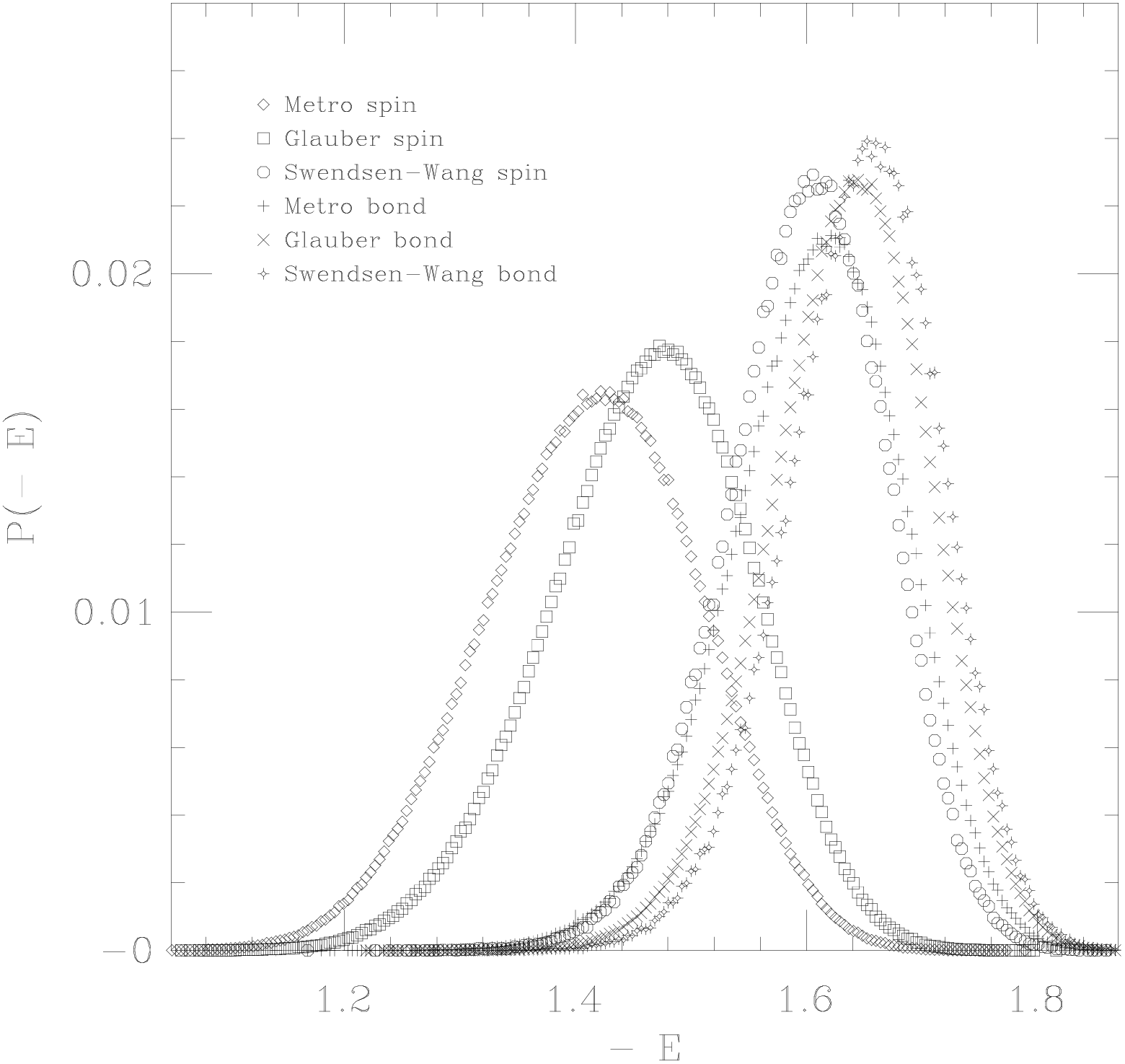}
{\bf Fig. 9b.-} Energy probability
distributions $P(-E)$ for the spin and bond versions of Metropolis,
Glauber and Swendsen-Wang dynamics at $\beta_1 = 0.35$, $\beta_2 =
0.6372$, $p=0.5$ and $L=32$ (critical point).

 \epsfxsize=330pt \epsfxsize=330pt \epsfbox{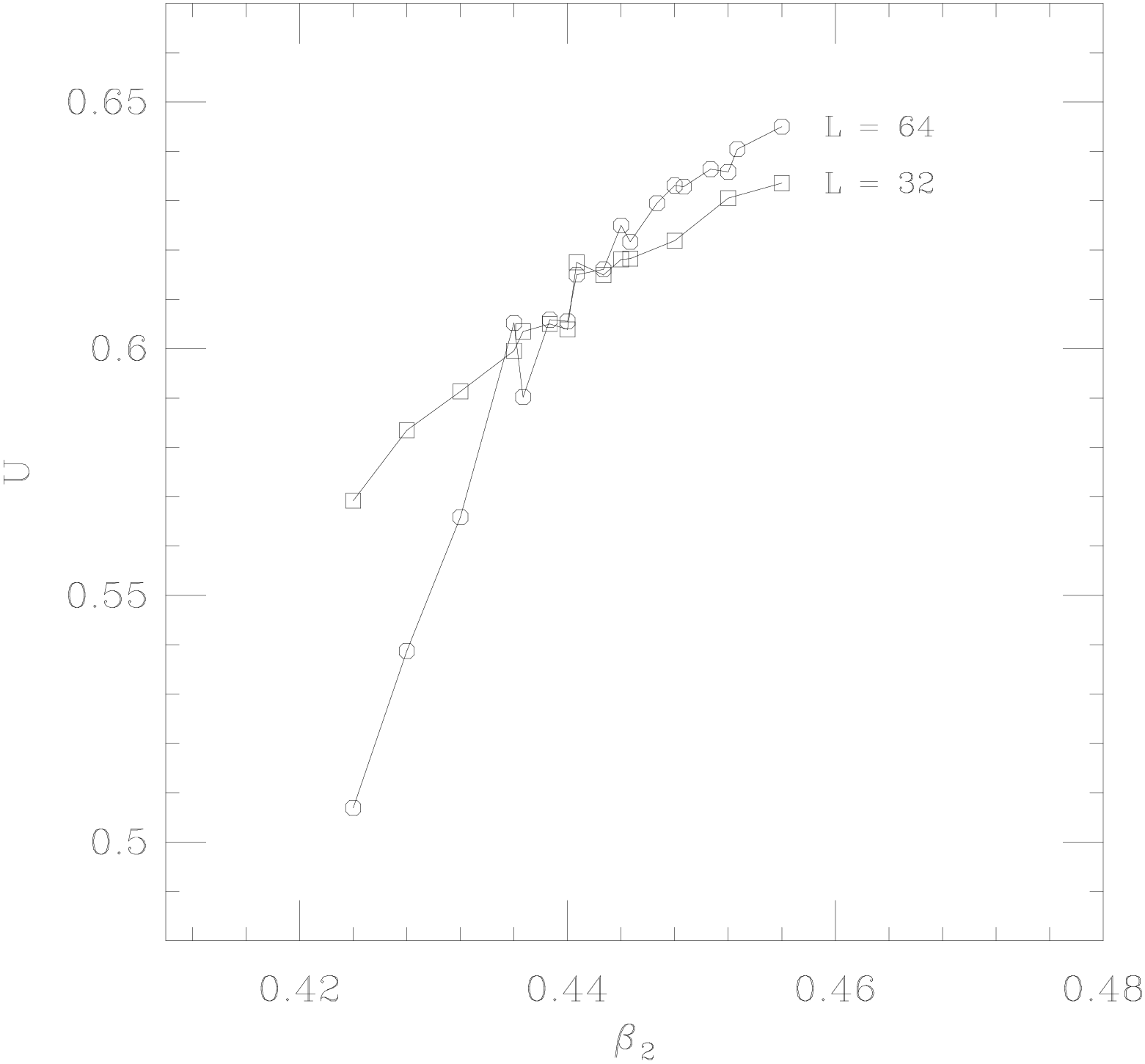}
{\bf Fig. 10.-} Cumulant values as a function
of $\beta_2$ (at $\beta_1 = 0.35$ and $\beta_3=0.6372$) and $L=32$ and
$64$  lattice   sizes  for   the  three-temperature   Metropolis  spin
dynamics.

 \epsfxsize=330pt \epsfxsize=330pt \epsfbox{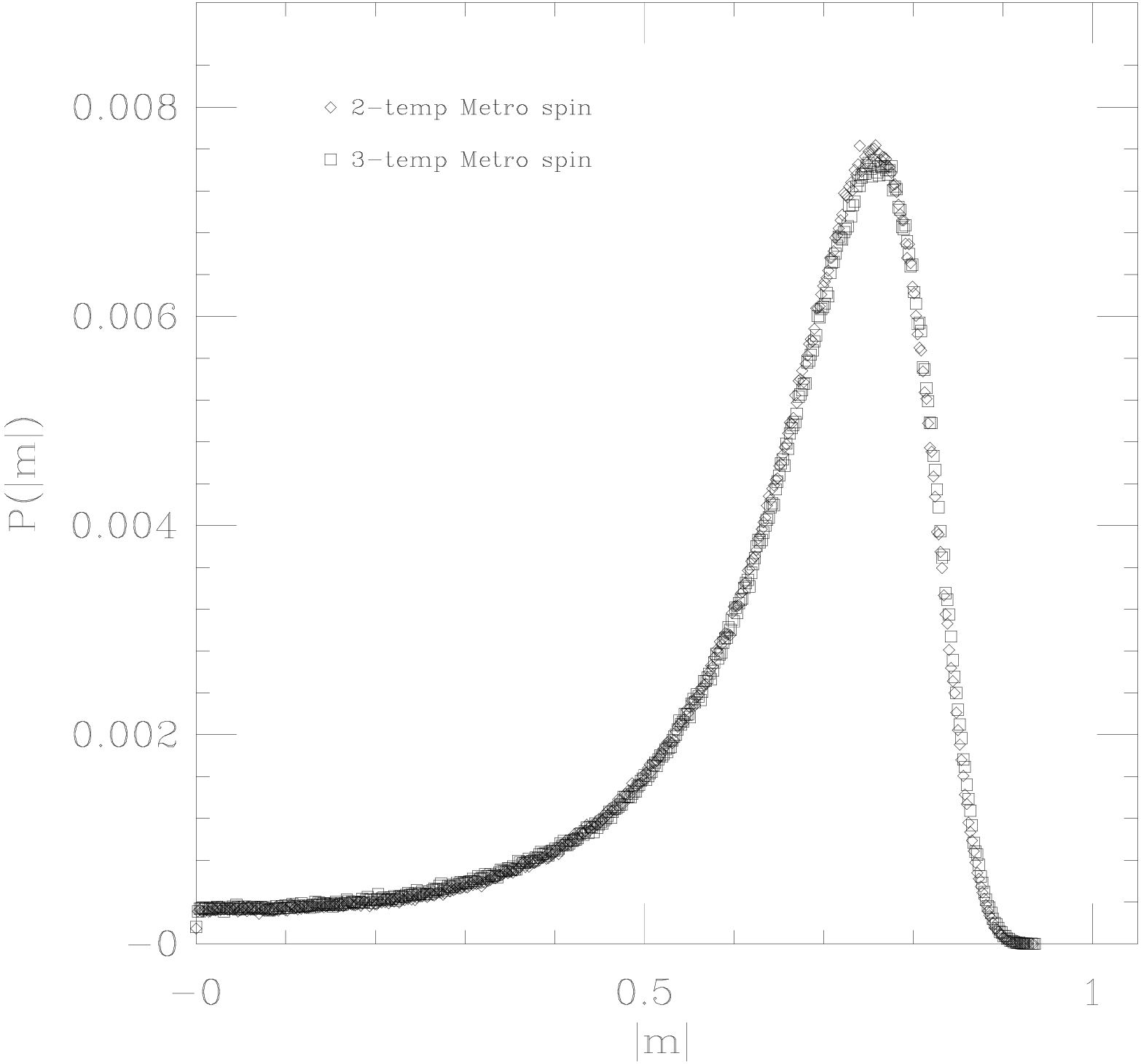}
{\bf Fig. 11a.-} Magnetization probability
distributions $P(|m|)$ for the two temperature and three temperature
Metropolis spin dynamics at the critical point ($L=32$).

 \epsfxsize=330pt \epsfxsize=330pt \epsfbox{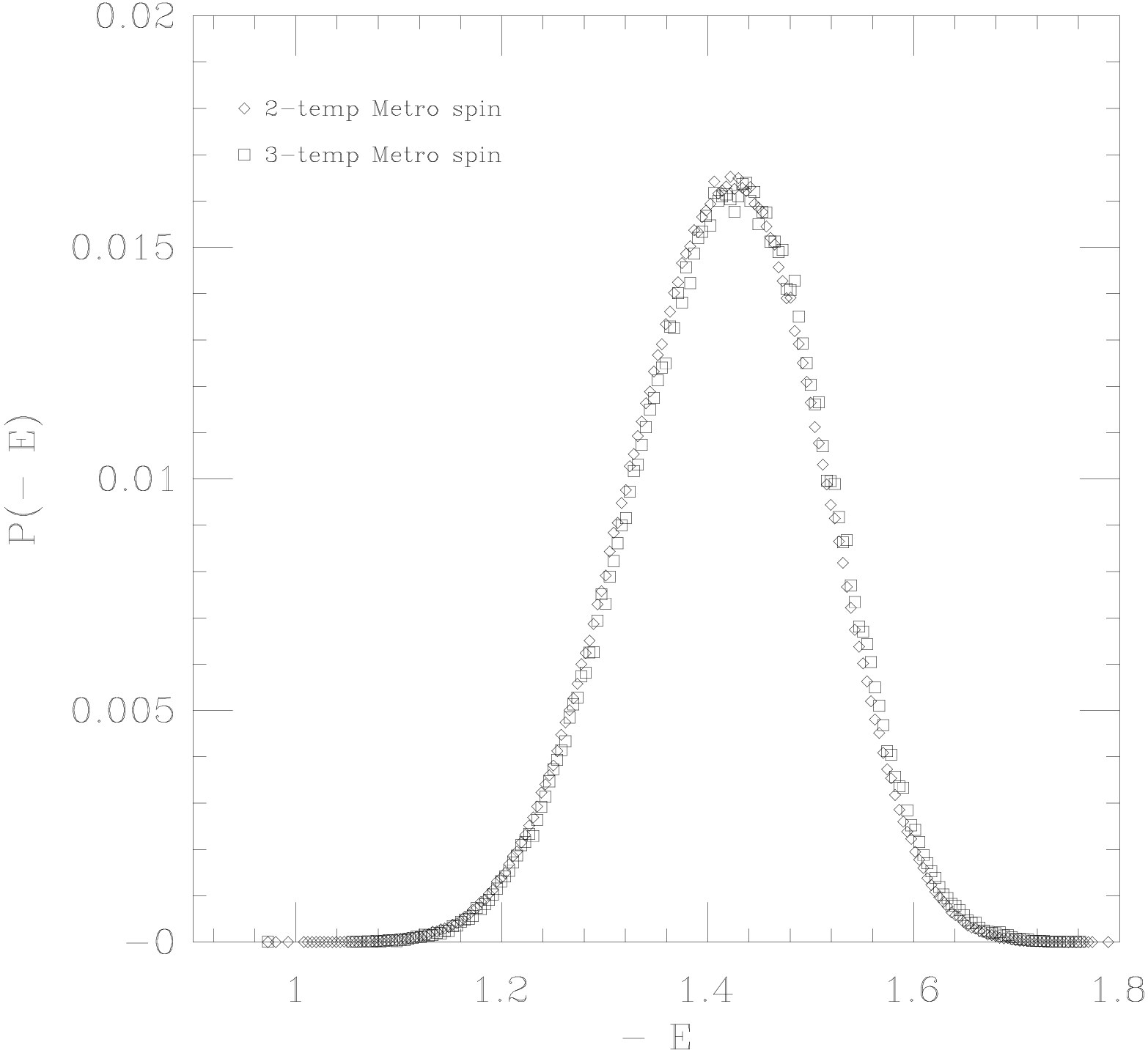}
{\bf Fig. 11b.-} Energy probability
distributions $P(-E)$ for the two temperature and three temperature
Metropolis spin dynamics at the critical point ($L=32$).

\end{document}